# General relativity and cosmology**

Martin Bucher[†,‡] and Wei-Tou Ni[*]

[†]*Laboratoire APC, Université Paris 7/CNRS,*
*Bâtiment Condorcet, Case 7020,*
*75205 Paris Cedex 13, France,* bucher@apc.univ-paris7.fr

[‡]*Astrophysics and Cosmology Research Unit and*
*School of Mathematics, Statistics and Computer Science,*
*University of KwaZulu-Natal,*
*Durban 4041, South Africa*

[*]*Center for Gravitation and Cosmology,*
*Department of Physics, National Tsing Hua University,*
*Hsinchu, Taiwan 30013 ROC* weitou@gmail.com

This year marks the hundredth anniversary of Einstein's 1915 landmark paper "Die Feldgleichungen der Gravitation" in which the field equations of general relativity were correctly formulated for the first time, thus rendering general relativity a complete theory. Over the subsequent hundred years physicists and astronomers have struggled with uncovering the consequences and applications of these equations. This contribution, which was written as an introduction to six chapters dealing with the connection between general relativity and cosmology that will appear in the two-volume book *One Hundred Years of General Relativity: From Genesis and Empirical Foundations to Gravitational Waves, Cosmology and Quantum Gravity*, endeavors to provide a historical overview of the connection between general relativity and cosmology, two areas whose development has been closely intertwined.

One hundred years ago the best model of the Universe summarizing the state of the observations at that time was the Kapteyn universe, which consists of a system of stars distributed more or less uniformly within a disk about 10 kpc in diameter and 2 kpc thick. In this model the Sun is situated near the center of the disk. In 1917 Einstein [1] proposed a static cosmological model based on the `cosmological principle,' a generalization of the Copernican principle postulating that the homogeneity and isotropy of space in the large should be extended to include the time dimension as well. Using distance determinations to about 100 globular clusters, Shapley in 1918 pushed back the boundaries of the measured Universe and concluded that the Sun lies near the edge of this distribution. As a result of rapid improvements in astronomical observations, the size of our observed universe soon grew to extend almost to our causal horizon.

This next group of chapters comprising Part IV of this GR100 book deals with Cosmology, applying Einstein's theory of general relativity to the universe as a whole. Cosmology considers the universe on very large scales and evolving over very long times, comparable to the age of the universe. Cosmology today is often described as being a `precision' science, and the insistence on this term reflects that cosmology had not always been seen as such. When Einstein formulated his general theory of relativity, little was known about the universe beyond our own galaxy, and although distant galaxies had been observed as `nebulous' unresolved spiral blotches, it was not at all clear that these `nebulae' consisted of numerous stars much like our own galaxy.

A glimpse of the state of affairs slightly after the formulation of general relativity may be gained by reading the written summaries [2] of the `Great Debate' held at the Smithsonian Institution in 1920, where Harlow Shapley and Heber Curtis sparred over the question of the `scale of the universe.' The former argued that the `nebulae' were simply clouds lying on the periphery of our own galaxy whereas the latter maintained that the observations at the time suggested that the `nebulae' were distant `island universes' much like our own galaxy. A key observation that helped settle this question in favor of the latter point of view was the discovery by Edwin Hubble of Cepheid variable stars in the Andromeda galaxy, which established that Andromeda lies far beyond the confines of our own galaxy.

On the observational side, as telescopes and other observational techniques greatly improved, our view of the universe progressively expanded to greater and greater distances. Once it had been established that the universe was populated by galaxies of different sizes, a recurrent question became whether the universe was homogeneous and isotropic on the largest observable scales, and it is only recently that galaxy surveys sufficiently deep and with sufficient statistics became available to settle this question definitively. An excellent historical account of these early debates and their role of the development of modern cosmology can be found in Peebles' 1980 book [3] (see also [4]).

The theory underlying the hot big bang model as we know it today was developed before the observational issues mentioned above had been settled. The geometry and time evolution of the universe as predicted by Einstein's theory are given by what is now known as the Friedmann-Lemaître-Robertson-Walker (FLRW) model, which describes the solutions to Einstein's field equations for a spatially homogeneous and isotropic universe whose scale factor varies with time. This solution to the Einstein field equations was first put forward by Alexander Friedmann in 1922 [5, 6] and later independently by Georges Lamaître [7] (see also [8]). Robertson [9] and Walker [10] subsequently showed that this was the only solution to the field equations consistent with spatial homogeneity and isotropy. As we discussed above, in 1917 Einstein [1] had put forth a theory of a static universe — a solution of the general relativity field equations that is not only homogeneous and isotropic in the three spatial dimensions but also homogeneous in time. Given the lack of compelling observational evidence to the contrary at the time, Einstein believed that an eternal universe, in which the Copernican principle held not only in three spatial dimensions but also in time, was more elegant and hence more plausible. In order to satisfy the gravitational field equations, Einstein had to introduce a cosmological constant term, denoted by $\Lambda$, a proposal that according to George Gamow, Einstein had later once described as his `greatest blunder,' although the authenticity of this quote is doubted by some.

Of the following six chapters of this volume dealing with the connection between cosmology and general relativity, the first three chapters deal primarily with the observations underpinning our modern conception of the cosmos. The first chapter of this group "Cosmic structure" provides a comprehensive overview of galaxy surveys and the information provided by them regarding the large-scale distribution of mass in the universe [11]. The second chapter discusses the physics of the 2.725 K cosmic microwave background (CMB) radiation [12]. The CMB confirms the big bang story in two ways. Firstly, the excellent agreement between the observed frequency spectrum of the microwave sky and a perfect blackbody spectrum at T=2.725 K provides strong confirmation of the expanding universe scenario, in which the universe at early times was much smaller and hotter, and thus once in a state extremely close to thermal equilibrium. Secondly, the measurement of the small departures from homogeneity and isotropy confirms the gravitational instability hypothesis for the origin of structure. This chapter describes in detail how the precise mapping of the anisotropies of the blackbody temperature in both intensity and polarization can be used to test theories of the very early universe based on new physics far beyond the standard model and how these fluctuations provide the initial conditions for the subsequent evolution leading to the formation of structure. The third chapter deals with the SN Ia distance scale and the discovery of the accelerated expansion of the universe [13]. The fourth chapter "Gravitational lensing in cosmology" details the present state of gravitational lensing probes [14]. The observation of the deflection of light rays as predicted by general relativity of course dates back to the famous expedition led by Arthur Eddington in 1919 to the island Principé off the west coast of Africa in order to measure the bending of light by the Sun during a total eclipse. Since that celebrated and crucial experiment confirming the theory of general relativity, gravitational lensing has evolved into a powerful tool for mapping the inhomogeneity in the distribution of mass.

As pointed out by Robert Dicke in the late 1960s [15, 16], a universe that is governed by general relativity (through the FLRW solution) and filled with ordinary matter (i.e., a combination of non-relativistic and relativistic particles with $w = p/\rho$ ranging from 0 to 1/3) provides a model that is unsatisfactory in several respects including: (1) the so-called `horizon problem,' whereby distant regions in opposite directions in the sky appear similar but could never have been in causal contact subsequent to the apparent initial singularity [17, 18], and (2) the `flatness problem,' under which the relation between the expansion rate and mean density at very early times usually expressed in terms of the dimensionless density parameter $\Omega(t) = 8\pi G\rho_{\mathrm{mean}}(t)/3H^2(t)$ initially had to be tuned near one with incredible precision to avoid a universe that today is either nearly empty or already re-collapsed [15, 16].

Although Dicke did not propose a concrete alternative, he suggested that the big bang model as formulated at the time might not be the whole story. "Cosmic inflation," reviewed in the fifth chapter [19], is a theory postulating that the very early universe underwent a period of quasi-exponential inflation during which whatever inhomogeneities may have existed prior to inflation are effectively erased and replaced with quantum fluctuations of the quantum fields relevant during inflation. The final chapter "Inflation, string theory and cosmic strings" [20] discusses possible connections between cosmic inflation and ideas from superstring theory concerning how inflation might be realized as part of a theory unifying all the fundamental interactions. Inflationary cosmology as developed in the early 1980s offered an attractive proposal, solving many of the problems that one would face if one naively extrapolated a conventional matter-radiation equation of state backward in time all the way to the putative big bang singularity. Moreover inflation offers a predictive mechanism for how the primordial perturbations from a perfectly homogeneous and isotropic FLRW background solution are generated. But inflation is an incomplete theory whose predictions are not completely defined without specifying a broader particle physics model within which inflation is realized. This final chapter reviews work on how inflation might be realized within the framework of superstring theory. These chapters serve to summarize the role of general relativity in modern cosmology.

Two key issues in modern cosmology are the dark matter problem and the dark energy problem. Since we have not included separate chapters for these issues, we give a brief review here. Both problems lead to postulating a new contribution to the stress-energy tensor because the known components combined with gravity as we understand it cannot account for the observations. In general, matter and objects in the universe have been discovered through non-gravitational means, for example through the electromagnetic radiation that they emit or scatter, but the history of objects first discovered through their gravitational pull on other objects dates quite far back, to the discovery of new outer planets.

Since Herschel's discovery of the planet Uranus in 1781, its observed orbit persistently deviated from its predicted Newtonian trajectory according to the ephemeris calculations at the time. Hussey suggested in 1834 that this disagreement could be explained by perturbations arising from an undiscovered planet. In 1846 Le Verrier predicted the position of this new planet. On September 25, 1846, Galle and d'Arrest discovered this new planet, now known as Neptune, to within a degree of Le Verrier's prediction. This discovery was a great triumph for Newton's gravitational theory and was the first example where deviations of an observed orbit from the predicted orbit led to the discovery of missing mass [21].

When sufficient data from meridian and transit observations of Mercury had been accumulated, in 1859 Le Verrier discovered a discrepancy between the observations and Newtonian gravity. This discrepancy may be described as the anomalous perihelion advance of Mercury,[1] at the rate of 38" per century. Using improved calculations and data sets, Newcomb in 1882 measured this discrepancy more precisely, obtaining 42".95 per century. A more recent value is (42".98 ± 0.04) per century [22].

In the last half of the 19th century, efforts to account for the anomalous perihelion advance of Mercury explored two general directions: (i) searching for a putative planet 'Vulcan' or other matter inside Mercury's orbit; and (ii) postulating an *ad hoc* modified gravitational force law. Both these directions proved unsuccessful. Proposed modifications of the gravitational law included Clairaut's force law (of the form $A/r^2 + B/r^4$), Hall's hypothesis (that the gravitational attraction is proportional to the inverse of distance to the $(2+\delta)$ power instead of the square), and velocity-dependent force laws. (The reader is referred to the book [23] for an in-depth history of the measurement and understanding of Mercury's perihelion advance.)

A compelling solution to this problem had to await the development of general relativity. When general relativity is taken as the correct theory for predicting corrections to Newton's theory, we understand why when the observations reached an accuracy of the order of 1" per century (transit observations), a discrepancy would be seen. Over a century, Mercury orbits around the Sun 400 times, amounting to a total angle of $5 \times 10^8$ arcsec. The fractional relativistic correction (perihelion advance anomaly) of Mercury's orbit is of order $\xi GM_{Sun}/dc^2$, (i.e., $8 \times 10^{-8}$) with $\xi = 3$ and $d$ being the distance of Mercury to the Sun. Therefore the relativistic correction for perihelion advance is about 40 arcsec per century. As the orbit determination of Mercury reached an accuracy of order $10^{-8}$, the relativistic corrections to Newtonian gravity became manifest.

We thus see how gravitational anomalies can lead either to the discovery of missing matter or to a modification of the fundamental theory for gravity. But there is also a third more mundane possibility. The secular contribution to the change of Moon's orbit owing to the back reaction of lunar tides of Earth is such an example.[2] When such orbital anomalies were found, their solution involved neither missing mass nor modified Newtonian dynamics. Moreover, the discovery of Pluto might be an example of an accidental discovery resulting from the interplay between theory and experiment. Observations of Neptune in the late nineteenth century made astronomers suspect that there could be another planet besides Neptune perturbing Uranus' orbit. In the early twentieth century, Lowell searched in vain for such a planet at the Lowell Observatory, which he founded in Flagstaff. In 1931, Tombaugh at Lowell Observatory discovered Pluto 6° off its predicted position. However its mass was much smaller than the predicted mass, so its influence on the orbits of Uranus and Neptune is negligible given the precision at that time. Pluto thus became the ninth major planet. However subsequently, as a result of a redefinition by the IAU of what constitutes a major planet, Pluto was downgraded to become reclassified as a `trans-Neptunian' object. In 1992 the next trans-Neptunian object was discovered, and since then more than 1200 such objects have been discovered with Eris (discovered in 2005) more massive than Pluto.

---

[1]Here `anomalous' means after the much larger corrections from perturbations of other planets have been subtracted. Only this residual, or anomalous part, constitutes a sign of something new.

[2]The action on the Earth is to slow down its rotation, so that after some time leap seconds need to be inserted.

We now turn to dark matter, which was first introduced to account for the dynamics of clusters and the rotation curves of spiral galaxies, because the ordinary visible baryonic matter in the form of stars assuming plausible mass-to-light ratios, and also of the gas in the case of galaxy clusters, was unable to account for the observations if the correctness of ordinary Newtonian gravity is assumed. (In this context corrections from special and general relativity are negligible.) The first hints of the need to postulate a dark matter component date back to the 1930s, when Zwicky found that the virial mass of galaxy clusters (i.e., the mass deduced by applying the virial theorem to the observed internal velocities of their member galaxies) greatly exceeds the total mass inferred from the luminous matter based on plausible mass-to-light ratios [24]. X-ray observations in the 1970s and 1980s alleviated this discrepancy without completely resolving it.

The need to postulate an additional dark matter component of some sort also arose from studies of the dynamics of spiral galaxies. In a series of papers in the 1970s, Rubin, Ford, and Thonnard measured the rotation curves of a number of disk galaxies and found that rotation speeds were larger than would be expected from the gravitational attraction arising from the visible mass distribution [25-27]. Moreover the shape of the rotation curves inferred from the visible mass did not agree with the observations, which found a rotation velocity almost constant with varying radius. The authors interpreted their findings as evidence for a new dark matter component.

Logically this conflict, known as the missing mass problem, could arise from a mass discrepancy, an acceleration discrepancy, or possibly even both. Many people believe that the missing mass is the dark matter, whose presence is made manifest only through its gravitational interaction with the visible baryonic matter. Others however explored the possibility that Newtonian gravitational dynamics should be modified. In 1982, Milgrom proposed the phenomenological MOND (MOdified Newtonian Dynamics) law for small accelerations [28]. Under this hypothesis, the gravitational dynamics become modified when the acceleration is smaller than $a_0 \sim 10^{-10}$ ms$^{-2}$ (Fig. 1). It was later shown possible to explain such a phenomenological Ansatz in the framework of a relativistic theory of gravity with additional degrees of freedom; however, none of these theories are yet satisfactory from a theoretical standpoint [31].

Perhaps today one of the strongest arguments for the need for a dark matter component arises from CMB anisotropy measurements as explained in the chapter on the CMB [12]. The standard six-parameter model, which includes a dark matter component comprising ~24% of the critical mass, provides an adequate fit to the observations, and it is not possible to account for the observations with a model having only baryonic matter. Compared to other probes, the CMB is a particularly clean probe of cosmological models and parameters because its interpretation is based on linear theory except for small and calculable nonlinear corrections. Except for a few degeneracies (which can be lifted by combining with a few other reliable ancillary data sets, such as BAO), the errors on the cosmological parameters as determined using the CMB are typically of order 1% and characterized by a well understood error budget. See [32, 33] for a detailed discussion of the current status of CMB constraints.

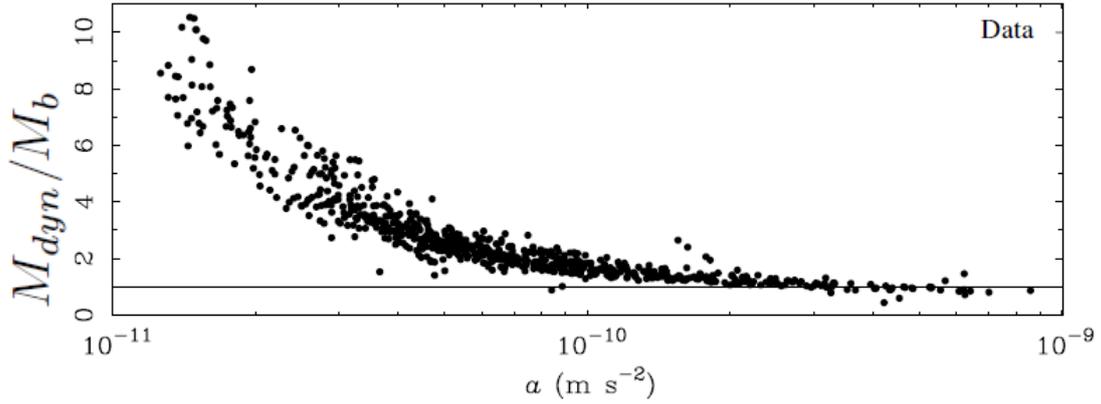

**Fig. 1. Critical acceleration from galactic rotation curves.** The ratio of dynamical to baryonic mass is shown at each point along rotation curves as a function of the centripetal acceleration at that point. The panel shows data for 74 galaxies [29]. The presence of missing or `dark' matter is the conventional explanation for why in spiral galaxies the measured rotation curves do not agree, neither in magnitude nor in shape, with the rotation curves predicted assuming a reasonable mass-to-light ratio and taking into account only the contribution from the disk. However, if the MOND hypothesis is adopted under which the gravitational dynamics becomes modified when the predicted acceleration is smaller than $a_0$ (~$10^{-10}$ ms$^{-2}$), the rotation curves can be fit assuming a constant mass-to-light ratio and no halo dark matter component. (Figure reprinted with permission from [30])

Another dramatic though less quantitative recent illustration of the need for dark matter arises from analyzing the bullet cluster as observed by gravitational lensing (which provides a clean probe of the mass) and as observed in X-rays. The bullet cluster, shown in Fig. 2 [34], is in fact the merger of two galaxy clusters. The X-ray image highlights the intracluster gas, which has been shocked and thus heated up as the result of the collision. This intracluster gas provides the dominant contribution to the baryonic mass, the mass from stars being subdominant. If there were no dark matter, the mass of the merged system would be concentrated in the center of the collision and would roughly coincide with the X-ray hot spot. But this is not what is observed in the gravitational lensing reconstruction of the projected mass density, which includes all types of mass whether visible or not. Instead the lensing reconstruction shows two halos which have hardly been disrupted by the collision apparently simply having gone through each other, as one might expect from a weakly interacting dark matter component. While striking, this special system should not be over-interpreted, because it is difficult to render quantitative this qualitative interpretation.

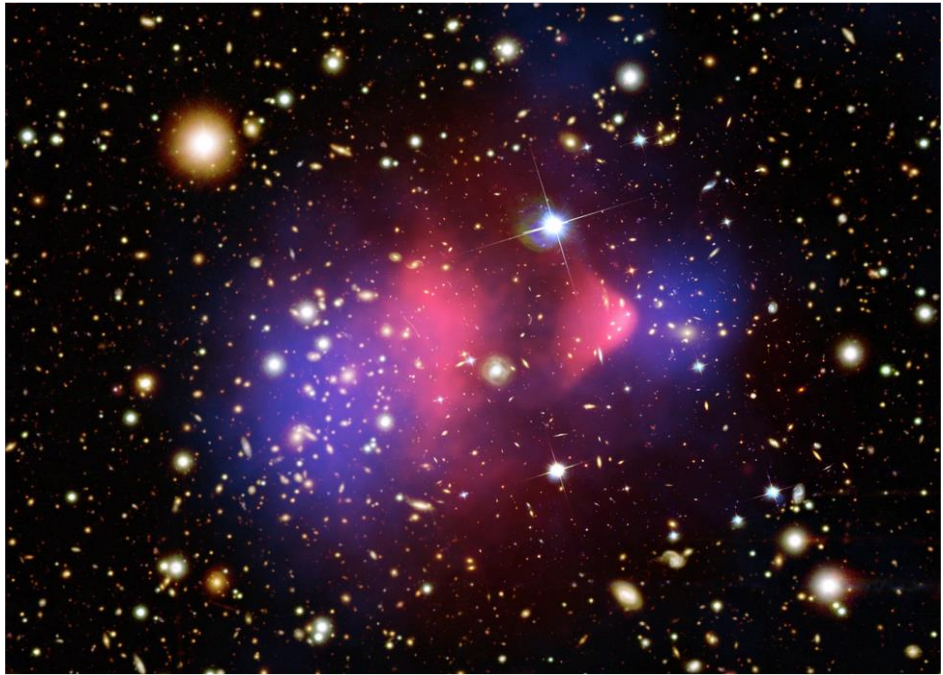

**Fig. 2. Chandra X-ray image of the "bullet cluster" (red) superposed with lensing image (blue).** In this merger seen in the X ray we see the hot baryonic intracluster gas that was shocked at the center of the collision. However in the gravitational lensing reconstruction we see two presumably dark matter halos comprising the majority of the mass which have been able to pass through each other without interaction. (*Credit*: *NASA, see* [34] for detailed explanation)

As the case for a dark matter component strengthened, more and more experiments to search for and ultimately characterize the dark matter have come on line. These experiments fall into two classes: direct and indirect. The direct dark matter search experiments look for the recoil of nuclei in the detector due to scattering of nuclei by dark matter particles. Backgrounds, for example from cosmic ray muons or from radioactive decays within or near the detector, obviously constitute a formidable obstacle. Indirect dark matter experiments search for an overproduction (or anomaly) of particles of various kinds from dark matter decays or annihilations. The current status of direct dark matter searches is summarized in Fig. 3 [35]. Indirect dark matter search experiments or observations may be subject to conflicting interpretations. The data from the ATIC and PAMELA experiments, for example, of the electron and positron fluxes have been interpreted as evidence for dark matter annihilation, although other explanations have been put forth. See the review [36] and references therein for a recent discussion of the methodology and current status of indirect searches. To date no confirmed dark matter candidates have been found. However, these experiments have made considerable progress in constraining models and in demonstrating and perfecting experimental techniques. Search experiments for example have already excluded a large part of the cMSSM-preLHC model parameter space.

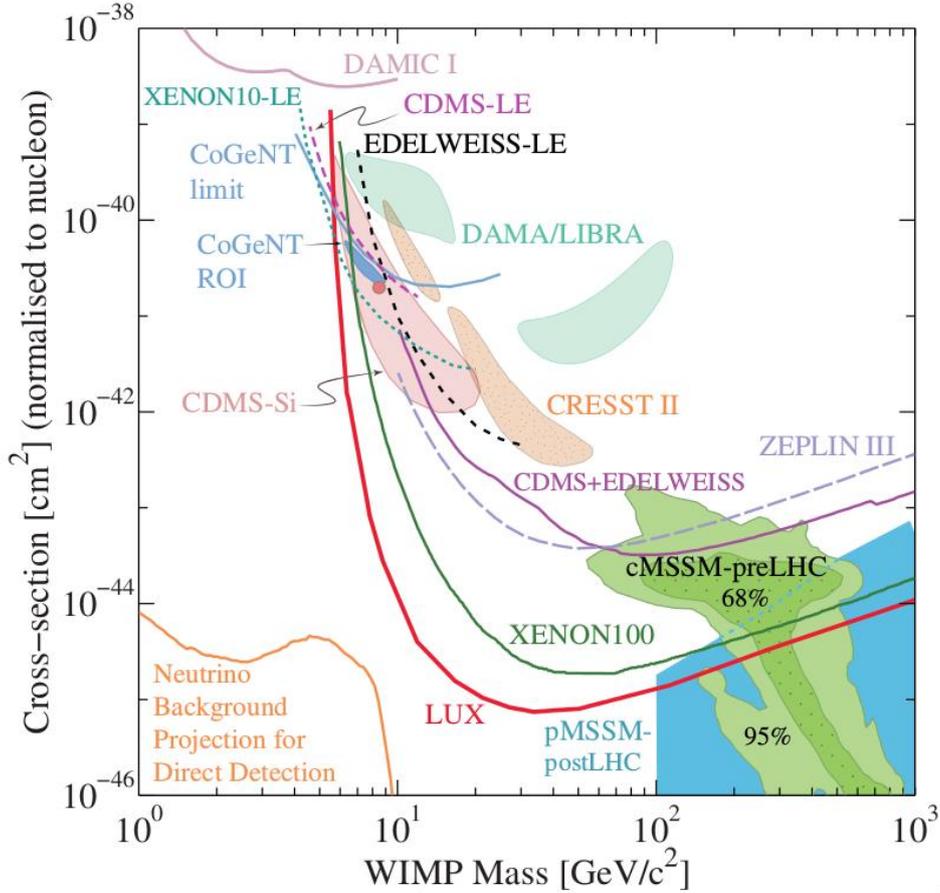

**Fig. 3. Particle dark matter searches: current status of constraints on WIMP dark matter from direct detection.** Here spin-independent couplings have been assumed. (Figure reprinted with permission from [35])

    The discovery of flat spiral galaxy rotation curves and the subsequent dark matter search takes us to the early 1990s when the favored cosmological model includes radiation (contributing negligibly to the mean density today), ordinary matter, and a weakly interacting dark matter component with a contribution such that the sum of the components yields a spatially flat universe. Despite the apparent beauty of a spatially flat universe with a vanishing cosmological constant, there were a few wrinkles to this story given the evidence at that time, including: (1) the inconsistency between the high measured value of the Hubble constant and the ages of the oldest known objects in the Universe, (2) the inconsistency between cluster baryon fraction and nucleosynthesis predictions, and (3) the inconsistency with the value of $\Omega_{matter}$ inferred from large-scale flows. (See, for example, [37, 38] and references therein.) This model was simpler than the six-parameter minimal model presently in vogue, and by the end of the 1990s it became clear that another component was needed to explain the current acceleration of the universe. This could be either a cosmological constant, or some other form of stress-energy with a similar equation of state (i.e., a component with a large negative pressure) that subsequently became known as `dark energy.'

    In 1970s when astrophysical and cosmological observational data accumulated, the improvement in the determination of the age of globular clusters [39, 40], the

Hubble constant, and the abundance of elements led to tensions in fitting FLRW cosmological models using only visible and dark matter. There were arguments that these measurements might already indicate a nonzero cosmological constant; however, the evidence was not yet compelling (see, e.g., [41-43]). In 1998 when the light curve correction to the intrinsic variability in absolute luminosities of the type Ia supernova had been perfected and a sufficiently large sample of such supernovae at intermediate redshift had been accumulated, it was shown that the expansion of the Universe was accelerating. Acceleration of the scale factor was inconsistent with the minimal $\Omega_{matter}$ = 1 cosmological model in vogue at the time and could be explained by a nonzero cosmological constant or other equally radical extension of the then accepted cosmological model. See the chapter *SNe Ia as a cosmological probe* for a detailed account of this development [13].

A key challenge of contemporary observational cosmology is to characterize the nature of the dark energy. If we ignore how the dark energy responds to cosmological perturbations, the problem can be expressed as measuring the run of the dimensionless parameter $w = p/\rho$ through cosmic history, where $p$ is the pressure and $\rho$ is the density of the dark energy. If the dark energy were a cosmological constant, it follows that $w = -1$ at all times, but a host of other scenarios have been proposed where $w$ is not exactly $-1$. Observations of the cosmic microwave background are ill-suited to characterizing the nature of the dark energy because the effect of the dark energy on the CMB anisotropies can for the most part be encapsulated into a single number — the angular diameter distance to the surface of last scatter of the CMB photons at z ≈ 1100. What is needed is a survey of the geometry of the universe out to large redshifts. From cosmological observations, it has been inferred that our universe is very close to being spatially flat. In a Friedman-Lemaître-Robertson-Walker (FLRW) universe, the luminosity distance is given by an integral functional of redshift and $w$:

$$d_L(z) = (1+z)(H_0)^{-1} \int_0^z dz' [\Omega_m(1+z')^3 + \Omega_{DE}(1+z')^{3(1+w)}]^{-\frac{1}{2}},$$

where $\Omega_{DE}$ is the present dark energy density parameter and the equation of state of the dark energy $w$ is assumed to be constant. For non-constant $w$ or non-flat FLRW universe, the expression is similar but slightly more complicated.

To determine $w(z)$, luminosity distances and angular diameter distances need to be measured at many redshifts where dark energy plays a role. These can be measured for example by using: (i) type Ia supernovae (as standard candles); (ii) baryon acoustic oscillations (BAO) in the matter power spectrum (as standard rulers); (iii) gamma ray bursts (as standards candle, although at present this method remains speculative); (iv) gravitational waves (GW) from compact binaries and SMBHs (as standard `sirens' or candles). All these methods suffer from dispersion and bias due to gravitational lensing of distant objects, which can substantially alter their apparent luminosities and sizes. GW methods have the potential for high precision, but one will likely have to wait about 20 years for a space-based GW observatory (e.g., eLISA [44]) to be put in place (see chapter 12 for a review on GW mission proposal studies [45]). GW methods are also limited by gravitational lensing.

However since the intrinsic measurement uncertainty can be made very small (better than 0.1 % with enough events), lensing uncertainty can be reduced by accumulating enough events for a detailed statistical analysis [46].

Assuming ΛCDM, we can investigate whether the parameter values found using the CMB observations are consistent with the parameter choices indicated by completely independent probes such as observations of type Ia supernovae and observations of the scale of the baryonic acoustic oscillations at various redshifts, as shown in Fig. 4. The fact that the three ellipsoids overlap indicates that ΛCDM is telling a consistent story.

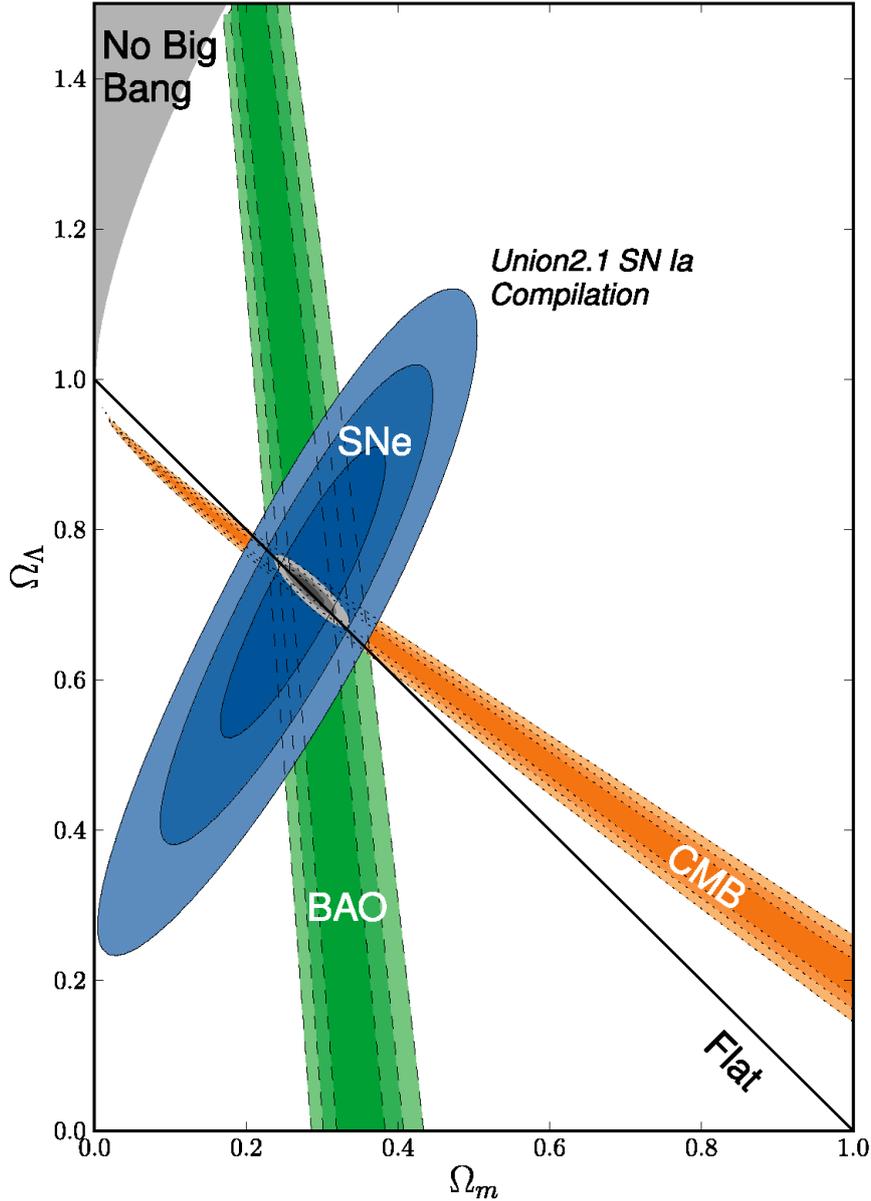

**Fig. 4. Concordance of CMB, BAO, and SNIa observations.** The respective error ellipsoids are shown on the $\Omega_m$-$\Omega_\Lambda$ plane. The supernova data are from the updated Union2.1 compilation of 580 Supernovae. (Figure reprinted with permission from [47].)

On smaller scales and at later times when nonlinear effects difficult to model are dominant, the success of the ΛCDM is less compelling. This is not because there is a clear contradiction between the predictions and what is observed but rather because the predictions of the theory are notoriously difficult to calculate because of the presence of nonlinear physics and poorly understood processes such as star formation and galaxy formation for which much of the modeling is at present still largely *ad hoc.* It has been argued by some that alternative models of gravity are able to account for the observations without positing an additional dark matter component. For example, in Fig. 1 it is argued that MOND theory offers a better explanation of galactic rotation curves. In MOND (Modified Newtonian Dynamics) proposed by Milgrom [28] in 1983 Newtonian theory remains valid for large accelerations, but becomes modified for centripetal accelerations smaller than the critical acceleration $a_0 \sim 10^{-10}$ ms$^{-2}$. Numerically, the critical acceleration is related to the cosmological constant Λ: $a_0 \sim \Lambda^{1/2}$ in natural units. Gravitational acceleration, of course, is not an invariant in Einstein's theory, so in order to incorporate MOND into general relativity some additional field would be required to define a preferred time direction, or equivalently a foliation of spacetime into 3-dimensional hypersurfaces. Stratified theories of gravity and preferred-frame theories of gravity introduce a preferred foliation. With a preferred foliation and extra degrees of freedom other than the physical metric, the theories still need to satisfy the strong equivalence principle to experimental precision to be viable. This is one reason why stratified theories or preferred-frame theories are not easy to construct in an empirically viable manner, so that they are consistent with experiments and observations on various non-cosmological-scales and with local experiments. Such a theory was constructed in 1973 [48, 49] and also constructed for MOND phenomenology in 2004 [31]. Nevertheless both these theories have encountered restrictions as empirical evidence is accumulated. For example, pulse timing observations on the relativistic pulsar–white dwarf binary PSR J1738+0333 provide stringent tests of these theories [50]. However, GR is covariant, but cosmology is stratified in the large. In studies of the microscopic origin of gravity and quantum gravity, the question arises whether Lorentz invariance is fundamental or derived, especially in the canonical formulation [51].

The histories of cosmology and relativity theory over the last 100 years have been closely intertwined. On the one hand, models of the universe rely on general relativity theory for part of their dynamics. On the other hand, the universe provides a testing ground for general relativity where Einstein's theory can be confronted with observation under the most extreme conditions: on the largest length scales, over the longest time intervals, and at the highest energy scales, close to the putative big-bang singularity. We close by identifying four areas where despite the existence of plausible hypotheses, it is likely that the last word has not been said and surprises may be forthcoming: (1) the dark matter-acceleration discrepancy problem; (2) the dark energy problem; (3) the inflationary epoch; and (4) a possible quantum gravity phase. It is likely that over the coming 100 years part of the story told here may be substantially rewritten.